\documentclass[runningheads]{llncs}
\usepackage[T1]{fontenc}
\usepackage{graphicx}
\usepackage{balance} 
\usepackage{amsmath}
\usepackage{amssymb}
\usepackage{mathtools}

\usepackage{float}
\usepackage{xr}   
\begin{document}

\title{An agentic system with reinforcement-learned subsystem improvements for parsing form-like documents}
\titlerunning{Accepted at AAMAS-EMAS}
%
\author{Ayesha Amjad\inst{1}\orcidID{0009-0002-4806-7207} \and
Saurav Sthapit\inst{2}\orcidID{0000-0002-7421-0479} \and
Tahir Qasim Syed\inst{1}\orcidID{0000-0003-0638-9689}}
\authorrunning{A. Amjad et al.}
%
\institute{Institute of Business Administration Karachi, Pakistan\\
\email{aamjad@khi.iba.edu.pk, tahirqsyed@gmail.com}\\
Coventry University, UK\\
\email{\{ae0066\}@coventry.ac.uk}}
\maketitle              
\begin{abstract}
Extracting alphanumeric data from form-like documents such as invoices, purchase orders, bills, and financial documents is often performed via vision (OCR) and learning algorithms or monolithic pipelines with limited potential for systemic improvements. We propose an agentic AI system that leverages Large Language Model (LLM) agents and a reinforcement learning (RL) driver agent to automate consistent, self-improving extraction under LLM inference uncertainty. Our work highlights the limitations of monolithic LLM-based extraction and introduces a modular, multi-agent framework with task-specific prompts and an RL policy of rewards and penalties to guide a meta-prompting agent to learn from past errors and improve prompt-based actor agents. This self-corrective adaptive system handles diverse documents, file formats, layouts, and LLMs, aiming to automate accurate information extraction without the need for human intervention. Results as reported on two benchmark datasets of SOIRE, and CORD, are promising for the agentic AI framework.

\end{abstract}
\section{Introduction}
A significant volume of form-like documents \cite{hu_question-answering_2023} is generated every day by several industry verticals \footnote{Industry verticals include retail, finance, healthcare, insurance, manufacturing, and businesses in general}, which includes invoices, purchase orders, receipts, bank statements, bills, and more. These documents require information extraction \cite{majumder_representation_2020} to serve downstream tasks and applications in structured formats, such as efficient archiving, fast indexing, and document analytics. However, both traditional and contemporary approaches to extracting structured information from these documents are limited in their capability to efficiently deal with varying layouts, formats, and complexity \cite{hu_question-answering_2023}.


Contemporary usage of deep learning algorithms such as convolutional neural networks (CNN) \cite{gilani_table_2017}, \cite{arslan_end_2022} and graph neural networks (GNN) \cite{riba_table_2019}, \cite{gemelli_graph_2022}, was especially influential for detecting tabular structure by leveraging its graphical properties. Methods proposing recurrent neural networks (RNN) \cite{rasmus_palm_cloudscan_2017}, and transformer architecture for key information extraction \cite{powalski_going_2021}, \cite{hong_bros_2021}, \cite{wang_lilt_2022}, \cite{hu_question-answering_2023} suggested spatial relationship, and semantic entity recognition as primary contributing factors. However, these models require large training data and boast complex architectures. Additionally, CNNs and GNNs focus primarily on the layout analysis to perform extraction, ignoring semantics and contextual awareness. Overall these models assume a list of predefined entities, hence, struggle with unseen documents and entities.

Recent studies have discovered remarkable propensity of human-like reasoning \cite{kojima2022large}, planning \cite{guo2024largelanguagemodelbased}, judgment, and rationality \cite{xu2024magicinvestigationlargelanguage} in LLMs, leading to the possibility of building applications with autonomous decision-making and execution called agents \cite{woodridge1995intelligent}. To solve the complex problem of automated data extraction effectively, a system of multiple LLM agents working in a collaborative environment while learning from past experiences is presented in this study. The evaluation within the framework is done through self-feedback systems \cite{madaan2023selfrefineiterativerefinementselffeedback} as well as metrics to compute matches and similarity between raw data and its extracted counterpart. Self-learning and optimization of each LLM agent is driven through Gymnasium \cite{towers2024Gymnasium}, a reinforcement learning framework, where action space is in a natural language setting.  

\subsection{Understanding Form-Like Documents}

In form-like documents, critical information resides as key-value pairs and line items. A Key-value pair is a pair of linked data items, where a key is the unique identifier to look up the value \cite{hu_question-answering_2023}, and is largely drawn from a small vocabulary of field-specific variants \cite{majumder_representation_2020}. Line items are a list of repeated instances of items typically detailing transactional information \cite{katti_chargrid_2018} that may or may not have graphical borders. 

Variations in these data artifacts are found across different document types. For example, a key phrase “Invoice\#” can also be written as “Invoice Number” or “Invoice No.” and its value can consist of either all digits or an alpha-numeric sequence. Moreover, a key phrase and corresponding value can be positioned relative to each other, e.g., a value can be found in the same line as the key phrase with an in-between separator such as a colon (:), or directly underneath. Certain key fields such as “Date” are found in all form-like documents, but most of the other fields are unique to the document type.

Traditionally, detecting line items on a text file has been a complex problem due to variations in its layout and the inability of models to “classify” a structure as tabular based on its content alone \cite{katti_chargrid_2018}. Computer vision algorithms using Hough transform \cite{arslan_end_2022} have been successful in detecting graphical borders, but challenges persist for tables without borders.

\section{Related Work}

Convolutional neural networks such as Fast R-CNN with Region Proposal Networks \cite{gilani_table_2017}, and encoder-decoder architectures like Chargrid \cite{katti_chargrid_2018} were effective for table and line items detection but struggled with ambiguous or boundary-less table layouts. \cite{arslan_end_2022} proposed YOLOv5 for table detection and predefined keyword searches for key-value extraction. Graph-based approaches enhanced performance by encoding spatial and semantic relationships; for instance, GNN-enriched node embeddings differentiate between tables, headers, and plain text in \cite{gemelli_graph_2022}. RNN-based systems such as CloudScan \cite{palm_cloudscan_2017} aided key-value extraction for invoices but were limited by their sequential processing and difficulty in generalizing to unseen layouts. Advances in representation learning \cite{majumder_representation_2020} and contrastive learning, particularly in FormNetV2 \cite{lee_formnetv2_2023}, improved adaptability by encoding neighboring context and pretraining on form structure, although they still assumed predefined entities.

Transformer-based and LLM-driven methods highlights the efficacy for layout- and language-agnostic information extraction. Layout-aware transformers such as XYLayoutLM \cite{gu_xylayoutlm_2022}, spatial transformers like BROS \cite{hong_bros_2021}, and semantic-rich models like TILT \cite{powalski_going_2021} and LiLT \cite{wang_lilt_202cao_extracting_2021} significantly outperformed earlier deep learning models. LiLT, in particular, enabled multilingual document parsing. GenIE \cite{josifoski_genie_2022} autoregressively generates structured triplets (subject, relation, object), while KPVFormer \cite{hu_question-answering_2023} used a Q\&A-based encoder-decoder architecture to extract key-value pairs. LLM-based frameworks offer a paradigm shift—methods such as trigger-based zero-shot extraction \cite{cao_extracting_2021}, instruction-tuned LLaMA via LoRA \cite{jiao_instruct_2023}, and multimodal fine-tuned GenIK \cite{cao_genkie_2023} demonstrate state-of-the-art flexibility, usability, and robustness, especially in low-resource or OCR-imperfect settings.

Existing models are trained on monolingual documents, except LiLT \cite{wang_lilt_202cao_extracting_2021}, finetuned for specific document categories like invoices, or assume a predefined list of entities. This limits their effectiveness with unseen documents and dynamic layouts. Their complex architectures and sensitivity to hyperparameters also hinder widespread adoption. Furthermore, standardized evaluation metrics to report the extraction accuracy and system confidence is essential for all probabilistic models in automated environments.

\section{Methodology}
\begin{figure*}[ht]
    \centering
    \includegraphics[width=1.0\textwidth]{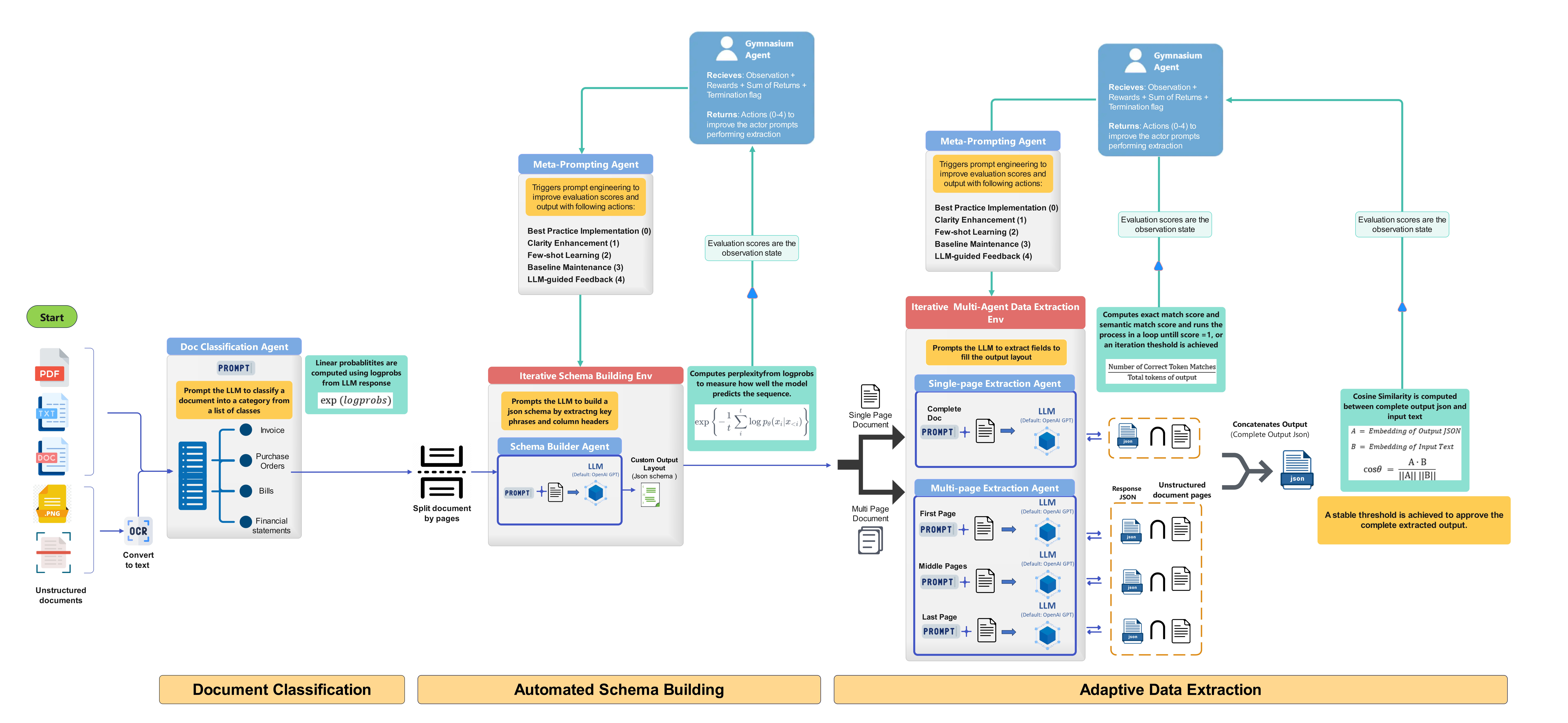}
    \caption{Architecture of the agentic form-like document data extraction framework comprising seven agents (blue), two Gymnasium environments (dark pink), and five evaluation metrics (cyan). Agents include a document classifier, schema builder, data extractor, two Gymnasium agents, a meta-prompting agent, and an evaluator. Evaluation metrics guide iterative optimization of the environments.}
    \label{fig:agentic_extraction}
\end{figure*}

The implemented framework (Figure~\ref{fig:agentic_extraction}) introduces a novel agentic approach to automated document data extraction by combining reinforcement learning with LLMs. As detailed in Section~\ref{sec:agentic}, this framework is evaluated against a baseline using both proprietary data and public benchmarks, including CORD \cite{park_cord_2019} and ICDAR-SOIRE \cite{huang_icdar2019_2019}. GPT-4o-mini \cite{openai2024gpt4ocard} serves as the default LLM, and LLaMA 3.3-70B \cite{grattafiori2024llama3herdmodels} is an alternative for processing sensitive or confidential documents. The token limit adheres to each model’s maximum capacity to handle lengthy inputs, and all responses are constrained to a strict JSON format for structured output.

\subsection{Document Text Reader}
A document reader class reads text off of form-like documents in various formats (PDF, TXT, DOC, PNG, JPG, JPEG, TIFF, BMP), while preserving the layout of the original document. For scanned PDFs and images, it uses state-of-the-art paddle OCR \cite{paddleocr2025}, ensuring accurate conversion into machine-readable text along with a built-in confidence score. Text-based formats (PDF, TXT, DOC) bypass OCR for downstream processing.

To preserve the original layout of documents post-OCR, we first compute the center position of each text block using \( x_{\text{center}} = \frac{\min(x_1, x_2, x_3, x_4) + \max(x_1, x_2, x_3, x_4)}{2} \). Layout preservation is then guided by a vertical threshold \( \Delta y_{\text{threshold}} = h_{\text{page}} \cdot 0.015 \), which helps identify line breaks and group related text blocks. To maintain appropriate spacing between elements, we calculate discrete block spacing as \( \max\left(1, \left\lfloor \frac{x_{\text{current}} - x_{\text{last}}}{w_{\text{page}} \cdot 0.01} \right\rfloor \right) \). These heuristics enable the reconstruction of structured layouts while accommodating document variability.

\subsection{Evaluation Metrics}
Extraction accuracy is evaluated using two non-parametric scores - exact match and cosine similarity. JSON output is corroborated against its ground-truth for exact matches, calculated as a simple ratio of intersection to union. It serves as a proxy for accuracy, helping validate extracted values and identifying hallucination in an LLM-generated output. 

{\small
\[\text{Exact Match} = \frac{\text{number\_of\_exact\_matches}}{\text{total\_fields}}\]
}

Cosine similarity score is computed between the generated JSON (A) and ground-truth JSON (B). It is robust against partial matches, and tolerant to schema variability. 

{\small
\[
\text{Cosine Similarity} = \cos(\theta) = \frac{\mathbf{A} \cdot \mathbf{B}}{|\mathbf{A}| |\mathbf{B}|} = \frac{\sum_{i=1}^n A_i B_i}{\sqrt{\sum_{i=1}^n A_i^2} \sqrt{\sum_{i=1}^n B_i^2}}
\]
}

The agentic process generates JSON schema dynamically for each document, leading to deviation in field and header names from its ground-truth and resulting in zero exact-match scores. To address this, a semantic match metric—using LLM-based evaluation but same scoring formula as exact match—is introduced.

Semantic match includes two variants: a basic version with binary field-level comparison and simple score aggregation, and an enhanced version with row alignment and cell accuracy:

{\small
\[
\text{Overall Score} = f(\text{confidence\_score}, \text{table\_analysis}, \text{field\_analysis})
\]
}
{\small
\[
\text{Table Score} = \frac{\text{row\_alignment} + \text{header\_match} + \text{cell\_accuracy}}{3}
\]
}

Classical machine learning metrics of \textit{F1}, \textit{precision}, and \textit{recall} are used particularly for a comparative analysis on benchmark datasets against current state-of-the-art methods.

\subsection{Baseline Data Extraction Framework}
\label{sec:baseline}

The process begins by reading text off of a form-like document, and appending it with a static "system" prompt. A real-time API call is made to the LLM, similar to a sequence-to-sequence task of Q\&A \cite{hu_question-answering_2023}, with upto 3 retries triggered by a predefined evaluation thresholds. The “system” prompt provides chain-of-thought instructions to establish context-aware extraction, and “user” prompt contains the source document text. The user prompt contains metadata specifying the page number, and document type (e.g. invoice, purchase order, utility bills). This contextual information helps the LLM understand the structure and semantics of the document.

\section{Agentic Data Extraction Framework}
\label{sec:agentic}

The agentic framework is designed to adapt to any document type, format, and LLM. The system employs a Gymnasium-compatible environment that models the extraction process as a Markov Decision Process (MDP) \cite{10.5555/528623}. It comprises of five fundamental components: actor agents, Gymnasium  environments, observed evaluation space, meta-prompting \cite{goodman2023meta} action space (~\ref{sec:meta}) , and RL agents (~\ref{sec:rl}). These components are essential building blocks of the self-directed and self-corrected multi-agent extraction pipeline shown in figure \ref{fig:agentic_extraction}.

\subsection{Document Classification Agent}

Document classification initiates the analysis of the document's first page using an LLM to categorize it into predefined classes: Invoice, Purchase Order, Utility Bill, Receipt, Financial Document (e.g., bank statements, balance sheets, investment reports), Salary Slip, or Unknown (fallback).

The classification confidence is approximated through a probabilistic approach, where the LLM's logarithmic probabilities are transformed into linear probability by taking an exponential of logprobs as shown below:
    
{\small
\[P_{\text{linear}} = e^{\text{logprob}} \cdot 100\%\]
}
Documents either failing to meet minimum $P_linear$ thresholds or a known category are flagged as 'unknown' for a manual review, thereby maintaining system reliability and accuracy.

\subsection{Automated Schema Building}

The schema building stage implements an iterative interaction between five components of agentic framework. The custom Gymnasium environment supports an episodic learning framework for schema actor prompt with the following components:

\subsubsection{State Representation:} Two-dimensional continuous space $S_{s}$, representing perplexity score $p \in [0,\infty)$, and schema complexity $c\in[0,1]$. Perplexity scores are computed by exponentiating the negative of the average of the logprobs. It captures the uncertainty of the generated schema, with no upper bound, its lower values indicate higher schema quality. 

{\small\[\text{p} = \exp\left(-\frac{1}{n}\sum_{i=1}^n logprobs\right)\]}
Schema complexity is a computed score between 0 and 1. It measures the structural complexity, where scores closer to 0 indicate simpler, more maintainable schema.
    {\small\[\text{c} = \alpha N + \beta D + \gamma B + \delta R\]}

where, $N$ is the nesting depth, $D$ is data type diversity. $B$ is the branching factor, and $R$ is the reference complexity. The weights are constrained such that:
{\small\[\alpha + \beta + \gamma + \delta = 1\]}
where:
$\alpha = 0.4$, 
$\beta = 0.2$,
$\gamma = 0.2$, 
$\delta = 0.2$.

\subsubsection{Reward Function:} Reward is computed linearly. The reward function $R_i(s_t, a_t)$ for state $s_t$ and action $a_t$ at time horizon $t$ uses perplexity score $p_t$ and complexity score $c_t$. Total reward is a simple aggregation of individual step rewards $r_i$.

{\small\[R_i(s_t, a_t) = (p_{\text{best}} - p_t) + (c_{\text{best}} - c_t)\]}

\subsubsection{Action Space:} The action space $A$ consists of a discrete set of five prompt engineering strategies, $A$  = \{0, 1, 2, 3, 4\}, each optimizing the prompts as per instructions (see~\ref{sec:meta}).

\subsubsection{Termination Criteria:} The schema building process continues until either maximum steps are reached, or no improvements are observed in consecutive iterations.
{\small
\[\text{terminated} = \begin{cases}
\text{1} & \text{if steps} \geq \text{max\_steps} \\
\text{1} & \text{if non\_improvement\_count} \geq 2 \\
\text{0} & \text{otherwise}
\end{cases}\]
}    

The dual optimization approaches (perplexity and complexity) provide complementary perspectives on schema quality, maintaining consistent improvement trajectories, and ensuring robustness while maintaining adaptability.

\subsection{Iterative Data Extraction Environment}

The multi-agent framework adopts an iterative approach, enabling interactions between the data extractor agent, extraction environment, meta-prompting action space, and an RL policy for optimizing actor prompts. It dynamically refines extraction strategies based on self-feedback .

Multi-page extraction is supported via a worker pool (Section~\ref{sec:sys}), where pages are handled in parallel and results are concatenated into the final output JSON.

As part of the experiment, the framework implements three distinct environment variants with episodic learning:
\begin{enumerate}
    \item DataExtractionEnvBase: Basic implementation with simple multiplicative reward mechanisms and fixed thresholds-based termination conditions.
    \item DataExtractionEnvIterative: Enhanced version using performance plateaus $P(t)$ to implement adaptive termination $T(s,t)$, and improvement-based with bonus reward system $R(s_t,a_t,s'_t)$ .
    \item DataExtractionEnvStepCount: Step-limited version with explicit exploration or exploitation trade-offs with time-penalization.
\end{enumerate}

Following architecture discusses the second environment \textit{DataExtractionEnvIterative} as the best performing framework for agentic data extraction.

\subsubsection{State Representation:} The state space $S$ is defined as a continuous 3-dimensional vector. $S$ is tracked over time horizon $T$, with best performing metrics and corresponding best prompt and output. For any time step $t \in T$: 

{\small
\[t^* = \underset{t \in T}{\operatorname{arg\,max}}(s_{exact}^t + s_{semantic}^t + s_{similarity}^t)\]
}

{\small
\[\mathcal{B} = \left\{\max_{t \in T} s_{exact}^t, \max_{t \in T} s_{semantic}^t, \max_{t \in T} s_{similarity}^t, \omega^{t^*}, \pi^{t^*}\right\}\]
}
where, $s^t$ are the scores at time $t$, and $(\omega, \pi)$ represent the output-prompt pair.

\subsubsection{Reward Function:} A step function for a combined score $\sigma_t = s_{exact} + s_{semantic} + s_{similarity}$, and an improvement bonus coefficient $\beta = 0.1$

{\small
\[R(s_t,a_t,s'_t) = \begin{cases}
    \beta +  \sigma_t - \sigma_{t-1} & \text{if } \sigma_t > \mathcal{B} \\
    \sigma_t - \sigma_{t-1} & \text{otherwise}
\end{cases}\]
}

\subsubsection{Termination Criteria:}The iterative environment implements adaptive termination $[T(s,t)$ based on performance plateaus $P(t)$ achievement and non-improvement tracking:

{\small
    \[P(t) = \begin{cases}
    1 & \text{if } \sigma_t \leq \sigma_{t-1} \text{ for } k \text{ consecutive steps} \\
    0 & \text{otherwise}
\end{cases}\]
}

where, $k = 2$ (configurable non-improvement threshold)

{\small
\[T(s,t) = \begin{cases}
    1 & \text{if } P(t) = 1 \\
    1 & \text{if } \text{steps} \geq \text{max\_steps} \\
    1 & \text{if } \text{threshold\_conditions\_met}(s) \\
    0 & \text{otherwise}
\end{cases}\]
}

\subsection{Meta-Prompting Agent}
\label{sec:meta}
The meta-prompting agent represents a systematic approach to dynamic prompt engineering, whereby, using driver prompts to refine and optimize actor prompts. These meta-prompting strategies makes up the discrete action space $A \in \{0,1,2,3,4\}$:
{\small
\[A=A(s, p, t, o, g) = \begin{cases}
    {\text{$Bt(p)$}} & \text{if } a = 0 \\
    {\text{$C(p)$ }} & \text{if } a = 1 \\
    {\text{$F(p, t_k)$}} & \text{if } a = 2 \\
    {\text{$N(p)$}} & \text{if } a = 3 \\
    {\text{$R(p, F(p, o, g))$}} & \text{if } a = 4
\end{cases}\]
}
\begin{enumerate}
    \item Best Practice Strategy ($Bt(p)$): Optimize prompts by letting LLM decide and establish best prompt engineering practices. 
    
    \item Clarity Enhancement Strategy ($C(p)$): Applies process decomposition, ambiguity elimination, tone optimization, structural enhancement, and goal clarification.
    
    \item Few-Shot Learning Strategy ($F(p, t_k)$): Implements example-based learning through carefully curated demonstrations for each task type $t_k$. A maximum of 3 examples are included to prevent context overflow.

    \item Feedback-Refine Optimization Strategy ($R(p, F(p, o, g))$): Takes inspiration from LLM-as-a-judge \cite{Gu2024ASO} policy, where an LLM evaluates and critics for complex tasks. It is a 2-stage process, where first a comprehensive feedback is provided by observing actor prompt $p$, generated output $o$ and corresponding ground truth $g$. Followed by refinements in actor prompt to minimize differences between generated and target output.
    
    \item Preservation Strategy (No Change) $N(p)$: Used as a control experiment that maintains original actor prompt.

\end{enumerate}

\subsection{Reinforcement Learning Agent} 
\label{sec:rl}
The Gymnasium agent implements a language model-guided policy $\pi(a|s)$ that observes the current state vector, current total rewards, and a boolean value for task completion, analyzes it through an LLM and maps to a meta-prompting action. This creates a closed-loop optimization system where prompt improvements are guided by both immediate feedback and long-term performance metrics.
{\small
\[\pi(a|s): S \rightarrow P(A)\]
}
   where, $P(A)$ is a probability distribution over action space $A$.

The system maintains state tracking to maximize the expected cumulative reward $R$ for all iterative environments with minimum terminal step $T$:

{\small\[argmax(\text{E}[\sum_{t=0}^{T} \gamma^t R])\]}

{\small\[T = \min(T_{\text{max}}, \inf\{t: n_t \geq N\})\]}

where, $T_{\text{max}}$ is the maximum allowed steps,
$n_t$ is the number of consecutive non-improvements, and
$N$ is the non-improvement threshold.

Exploration rate decay $\epsilon(t)$ of Gymnasium RL policy states: 
{\small
\[\epsilon(t) = \epsilon_0 \cdot e^{-\delta t}\]
}
where,
$\epsilon_0 = 1.0$ (initial exploration rate), and 
$\delta = 0.2$ (decay rate)

\subsection{Learned Prompt Optimization} 
\label{sec:lpo}

Learned Prompt Optimization (LPO) extends the capability of a simple RL agent by leveraging contextual bandit learner to iteratively explore and exploit effective meta-prompting strategies. It comprises of three core components:
\begin{enumerate}
    \item Vector Embedding: Converts textual inputs into dense vector representations using OpenAI ada-002 embedding model.
    \item Selection Scorer: Evaluates prompt strategies based on context and historical performance.
    \item Policy Learner: Updates strategy selection weights using Vowpal Wabbit \cite{vowpalwabbit_rl} contextual bandit implementation.
\end{enumerate}

For each iteration $t$, the probability of selecting action $a_i\in \ A$ is:
{\small
\[
P(a_i | C_t) = \frac{\exp(\theta_i^T \phi(C_t))}{\sum_{j=1}^{n} \exp(\theta_j^T \phi(C_t))}
\]
}
where \( \phi(C_t) \) is the context embedding and \( \theta_i \) are learned parameters. After observing cumulative reward \( R \), parameters are updated as follows, where \( \eta \) is the learning rate.

\[
\theta_i \leftarrow \theta_i + \eta R \nabla_{\theta_i} \log P(a_i | C_t)
\]

\subsection{System Optimization Components} 
\label{sec:sys}

The agentic data extraction framework encounters operational challenges with multi-page documents. A single page requires time $t$ with multiple LLM calls for optimized output, hence, processing $n$ pages scales linearly to $nt$, leading to significant computational overhead. Additionally, concatenating text from all $n$ pages into a single LLM call, while seems feasible, dampens output accuracy.

To overcome these challenges and build a robust document processing system with minimal redundant operations, parallel processing and caching mechanisms are implemented. Furthermore, our sequential framework, where the output from each step influences subsequent steps, requires comprehensive logging and error handling for debugging and monitoring to maintain system reliability.

\section{Results}

Results are reported for 32665 form-like documents, including 20295 proprietary documents and 11000 benchmark files, out of which approximately 93\% requires an OCR processing. The largest volume of document type is invoice with 17492 files, followed by 12987 receipts. Most documents comprises of 2-3 pages ($\text{Average Number of Pages} = \frac{\text{Total Number of Pages}}{\text{Number of Files}}$ = 2.7).

\subsection{Baseline Framework Results}

Table \ref{tab:one-shot-prompt-results} reports the baseline results of the LLM-generated output on 32665 files, via one-shot prompting, on two metrics - exact match, and cosine similarity. All reported scores are best of 3 pooled responses averaged over \# of files for each document type. Two factors have been observed to influence the evaluation performance, 1) Number of pages, 2) Scan quality. 

{\small
\begin{table}[H]
\centering
\caption{Results from the one-shot single-execution prompt} 
\begin{tabular}{|l|l|l|l|l|l|}
\hline
\textbf{Document Type}    & \textbf{Formats}   & \textbf{\# of Files} & \textbf{Exact Match} & \textbf{Similarity}  \\ \hline
Invoices                  & Images             & 10,000               & 0.43         & 0.61               \\ \hline
Invoices (Australian)     & Digital PDF        & 2202                 & 0.58          & 0.74                \\ \hline
Invoices (Australian)     & Scanned PDF        & 5290                 & 0.33          & 0.38                 \\ \hline
Purchase Orders           & Digital PDF        & 19                   & 0.78          & 0.76               \\ \hline
Purchase Orders           & Scanned PDF        & 138                  & 0.43          & 0.55               \\ \hline
Receipts                  & Images             & 987                  & 0.78          & 0.81               \\ \hline
Financial documents       & Mixed/HTML         & 2,832                & 0.30          & 0.59               \\ \hline
Utility Bills             & Images             & 100                  & 0.31          & 0.75               \\ \hline
Salary Slips              & Images             & 97                   & 0.84          & 0.86               \\ \hline
CORD                      & Images             & 10,000               & 0.71         & 0.90              \\ \hline
ICDAR-SROIE                & Images             & 1,000                & 0.67         & 0.75                \\ \hline
\end{tabular}
\label{tab:one-shot-prompt-results}
\end{table}
}

Table \ref{tab:one-shot-prompt-benchmark-results} extends the analysis by reporting on classical machine learning metrics of $F1$ , $precision$, and $recall$ for benchmark datasets.

{\small
\begin{table}[H]
\centering
\caption{Results from baseline framework on two benchmark datasets}
\begin{tabular}{|l|l|l|l|l|l|}
\hline
\textbf{Document Type}  & \textbf{F1 Score} & \textbf{Precision}  & \textbf{Recall} \\ \hline
CORD                             & 0.886          & 0.902      & 0.871      \\ \hline
ICDAR-SROIE                       & 0.808          & 0.780      & 0.838        \\ \hline
\end{tabular}
\label{tab:one-shot-prompt-benchmark-results}
\end{table}
}

\begin{figure}[H]
    \centering
    \includegraphics[width=0.6\textwidth]{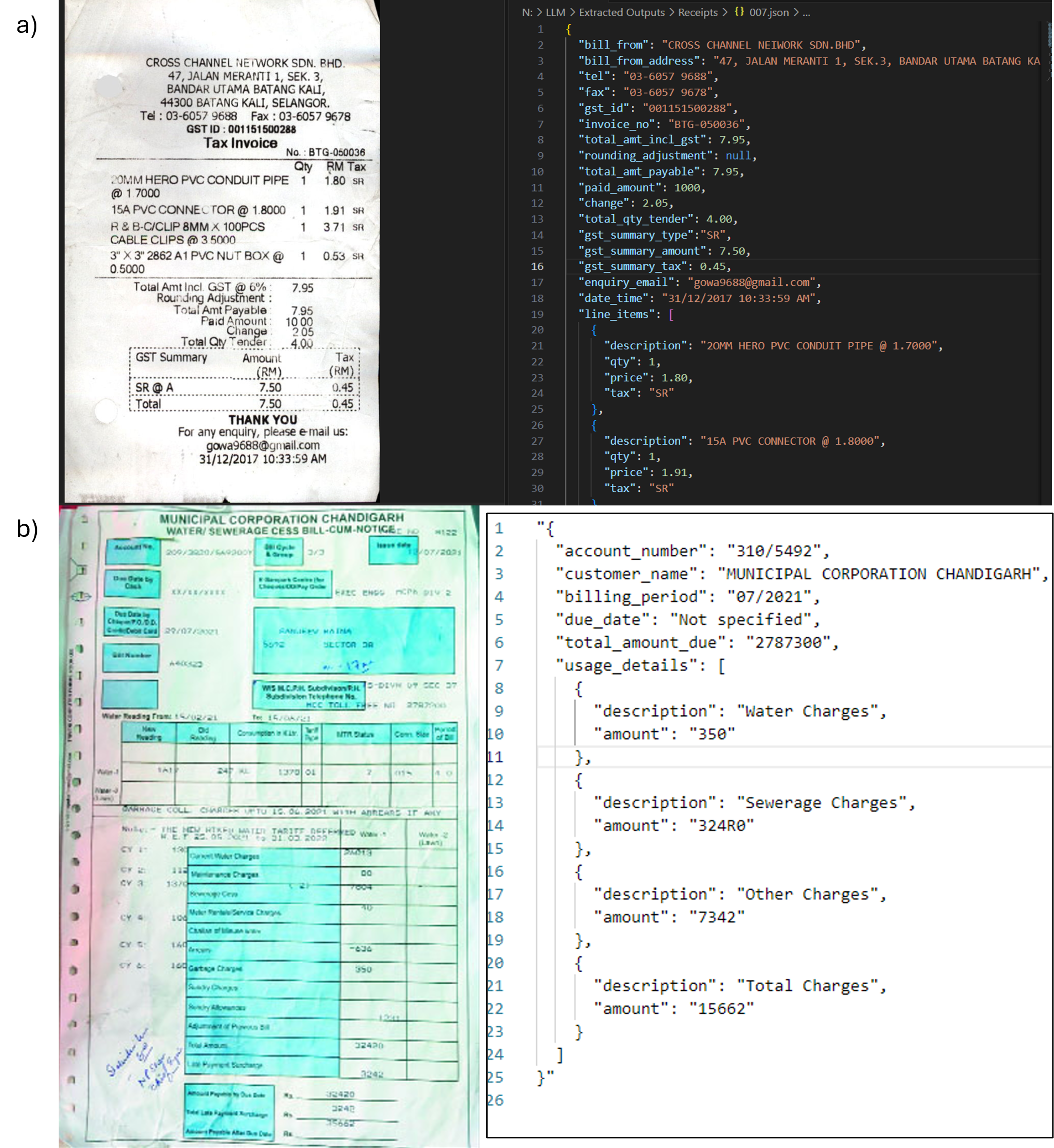}
    \caption{a) Extracted JSON of a receipt (96 dpi) scored 90\% on exact matches, 100\% on semantic matches, and 96\% on cosine similarity. b) Extracted JSON of a utility bill (66 dpi) scored lower: 30\% exact matches, 65\% semantic matches, 71\% similarity.}
    \label{fig:ocr}
\end{figure}

Financial documents, typically spanning 7 or more pages and available in high-quality formats, yield low extraction scores (30\% exact matches, 59\% similarity). In contrast, single-page images of receipts at 100 dpi (standard threshold = 300 dpi )—achieve higher scores (78\% exact, 81\% similarity). Similarly, single-page utility bills with even lower quality (89 dpi) still outperform financial documents, but falls behind receipts, in extraction accuracy (See figure \ref{fig:ocr}).

\begin{figure}[H]
    \centering
    \includegraphics[width=0.5\textwidth]{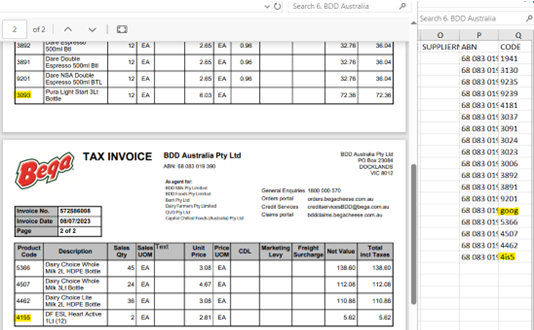}
    \caption{Incorrect data extracted for a 2-page scanned document. “3093” is extracted as “goog” and “4155” is extracted as “4is5” }
    \label{fig:incorrect}
\end{figure}

Two patterns are observed in the extracted JSON of multi-page documents; partial extraction across all pages (often missing line items from middle pages), and complete omission of entire pages (more common). Multiple reruns revealed inconsistent outputs, suggesting nondeterministic behavior from the LLM. The nondeterministic behavior worsens with poor quality scans, which sometimes returned a blank JSON or incorrect (See figure \ref{fig:incorrect}) and hallucinated response.

\begin{figure}[H]
    \centering
    \includegraphics[width=0.5\textwidth]{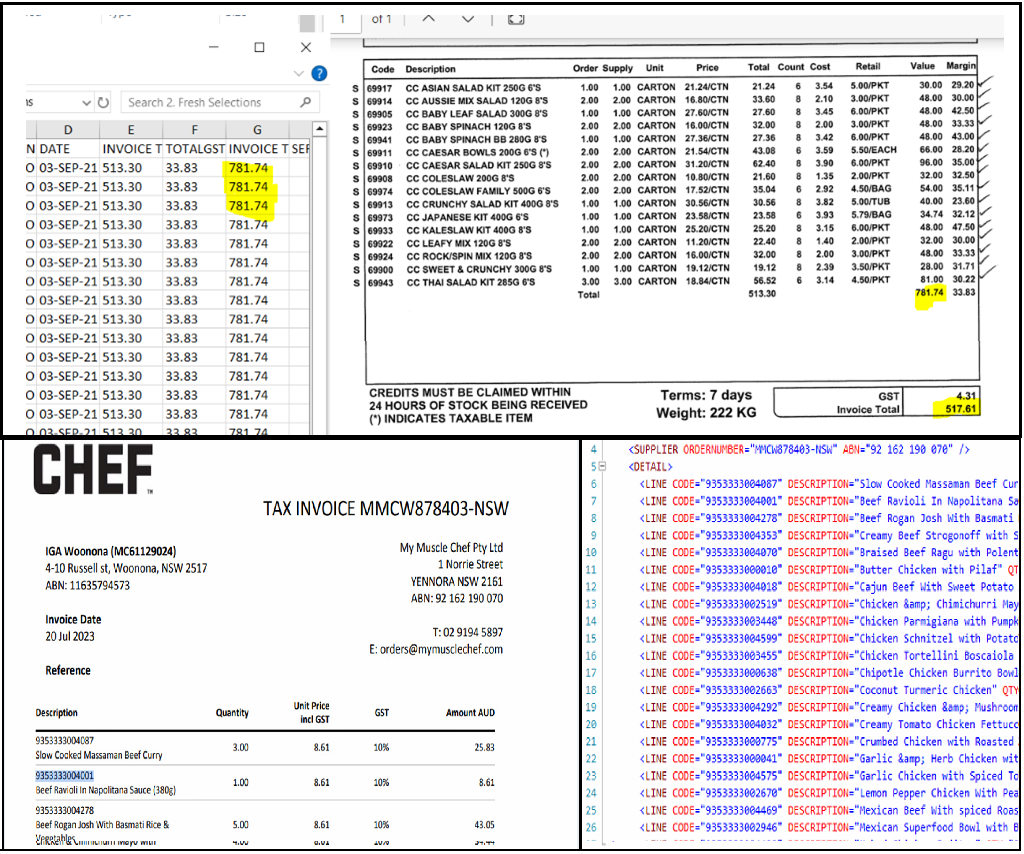}
    \caption{Top: 'Invoice Total' is misidentified as 'total' due to semantic closeness. Bottom: explicitly specifying 'CODE' as a column header enables correct extraction, even though there is no 'CODE' column header in the document table. }
    \label{fig:schema}
\end{figure}

Limitations of static prompts without explicit output schema are evident in cases with atypical key phrases or headers. Cases discussed in figure \ref{fig:schema} highlights the importance of including output schema in the extraction prompts. Extraction accuracy is influenced not just by page count and scan quality, but also by document formatting. Multi-line headers and dark-shaded key phrases often reduce accuracy, mainly due to issues in text conversion rather than LLM behavior. Processing documents as images generally yields more reliable results than pre-converting them to text.

The baseline monolithic prompt-based system exhibit nondeterministic behavior, producing inconsistent outputs with multi-page documents, tables with missing headers, poor quality scans, ambiguous field names, and atypical document formatting. Context window limitations further exacerbate the issue, restricting processing to shorter documents. There is also a higher likelihood of hallucination resulting from information density of multi-page documents.

\subsection{Agentic Framework Results}

Table \ref{tab:agentic data extraction results} reports the results of agentic data extraction framework on three evaluation metrics. The exact match scores are computed after a manual verification of each output JSON against its groundtruth, therefore, depicting the most accurate picture of the output performance. Semantic match is used to balance out a score of zero on the exact match during runtime iterative processing.

{\small
\begin{table}[H]
\centering
\caption{Results from the self-directed and self-corrected multi-agent system - \textit{gpt-4o-mini}} 
\begin{tabular}{|l|l|l|l|l|l|}
\hline
\textbf{Document Type}    & \textbf{Formats}   & \textbf{\# of Files} & \textbf{Exact Match} & \textbf{Semantic Match} & \textbf{Similarity}  \\ \hline
Invoices (mixed)          & Images             & 10,000               & 0.863         & 0.901      & 0.941         \\ \hline
Invoices (Australian)     & Searchable PDF        & 2202              & 0.985          & 0.944      & 0.908        \\ \hline
Invoices (Australian)     & Scanned PDF        & 5290                 & 0.913          & 0.900      & 0.939        \\ \hline
Purchase Orders           & Searchable PDF        & 19                & 0.994          & 1.00      & 0.980        \\ \hline
Purchase Orders           & Scanned PDF        & 138                  & 0.953          & 0.899      & 0.952        \\ \hline
Receipts                  & Images             & 987                  & 0.834          & 0.910       & 0.866        \\ \hline
Financial documents       & Images             & 2,832                & 0.962          & 0.998      & 0.927        \\ \hline
Utility Bills             & Images             & 100                  & 0.425          & 0.663      & 0.817         \\ \hline
Salary Slips              & Images             & 97                   & 1.00          & 0.998      & 0.959        \\ \hline
CORD                      & Images             & 10,000               & 0.866          & 0.812      & 0.921          \\ \hline
ICDAR-SROIE               & Images             & 1,000                & 0.753          & 0.911        & 0.890        \\ \hline
\end{tabular}
\label{tab:agentic data extraction results}
\end{table}
}

A clear improvement is observed in extraction completeness from every page of financial documents. For instance, if a bank statement page lists 15 transactions, all are accurately captured in the JSON, regardless of its page index or document length. This boosts the  exact match scores from 30\% to 96.2\% and similarity scores from 59\% to 92.7\%, primarily due to multi-page handling where each page is processed independently.

Searchable PDFs or high resolution single-page images achieve near-perfect extraction completeness and accuracy. Purchase Orders show a semantic score of 1.00, exact match of 0.994, and cosine similarity of 0.98. Similarly, each salary slip achieves a 100\% exact match and ranks only second to Purchase Orders on other metrics.

\begin{figure}[H]
    \centering
    \includegraphics[width=0.8\textwidth]{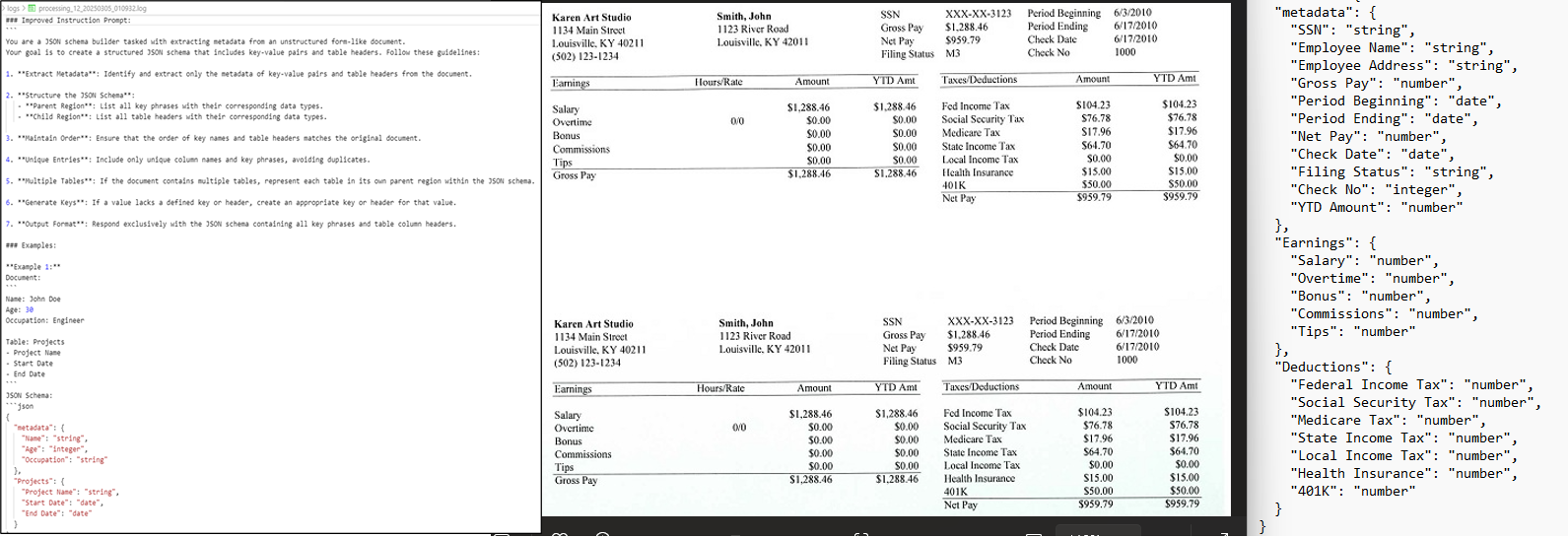}
    \caption{This salary slip shows four tables, but only two are unique, as earnings and tax deductions each appear twice. Schema builder prompt (left) retains only unique tables.}
    \label{fig:duplicate-tables}
\end{figure}

The agentic process remains robust against large file sizes and complex structure. When a document contain multiple tables, the algorithm extracts each as a distinct child collection in the JSON output without overly complicating the schema. As shown in figure \ref{fig:duplicate-tables}, even with duplicate tables, the schema builder agent filters out redundancies to retain only unique instances for extraction.

Low-quality scans ($\leq$ 100 dpi) remain challenging due to the framework's reliance on OCR, as evident by the poor performance of utility bills scoring lowest across all metrics. A potential solution is to bypass OCR and use multimodal LLMs for direct image-based data extraction.

\begin{figure}[H]
    \centering
    \includegraphics[width=0.6\textwidth]{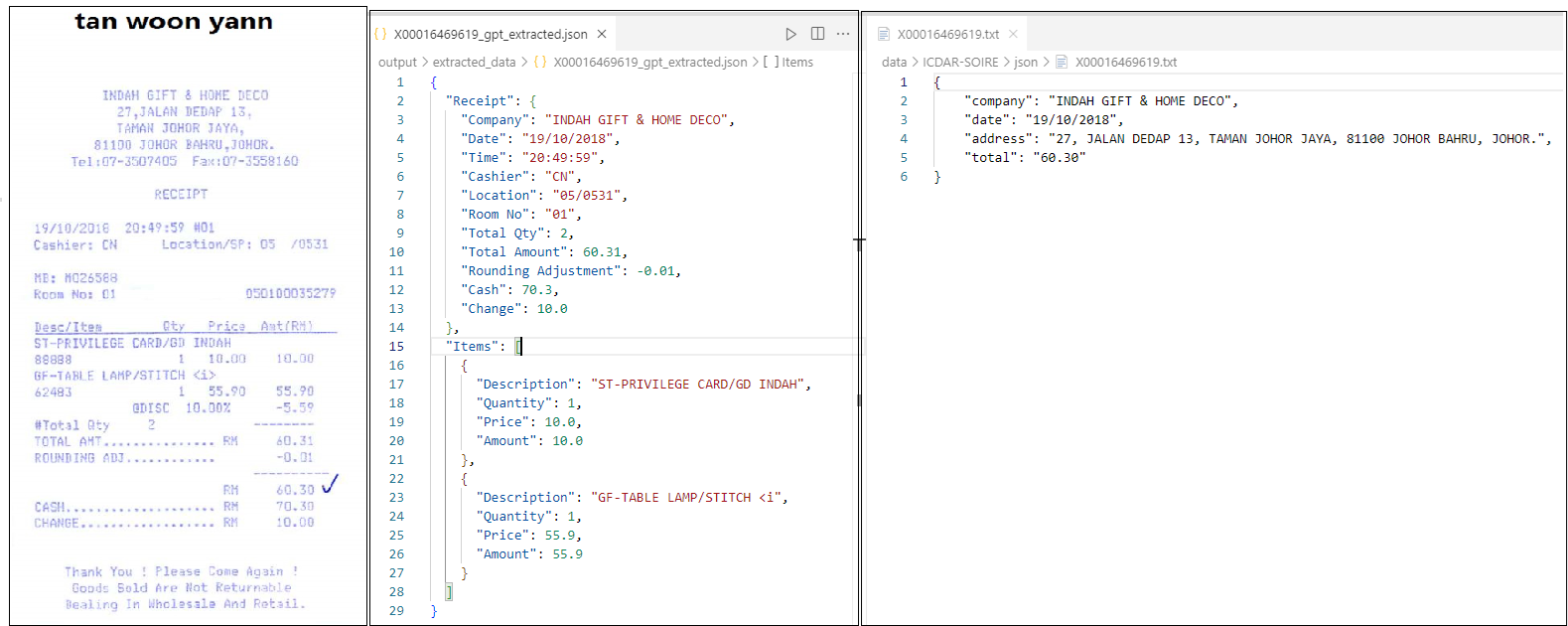}
    \caption{ICDAR-SROIE data collection (left), agentic extracted JSON (center), groundtruth (right)}
    \label{fig:icdar}
\end{figure}

The agentic process dynamically generates and optimizes the schema at runtime, enabling it to extract nearly all relevant fields—often beyond the scope of benchmark groundtruth. In figure \ref{fig:icdar}, the extracted JSON of a receipt includes all expected fields except one (address) and also captures additional fields not present in the groundtruth. To accommodate schema variations between the groundtruth and extracted output, a flexible evaluation using fuzzy matching is employed for \textit{F1}, \textit{precision}, and \textit{recall} (Table \ref{tab:agentic benckmark}). Values are considered a match if their similarity is $\geq$ 80\%, allowing minor formatting differences to be ignored.

\begin{table}[H]
\centering
\caption{Agentic results on benchmark shows a minimum 0.05 point boost on classical ML metrics}
\begin{tabular}{|l|l|l|l|l|l|}
\hline
\textbf{Document Type}  & \textbf{F1 Score} & \textbf{Precision}  & \textbf{Recall} \\ \hline
CORD                           & 0.965          & 0.966      & 0.964      \\ \hline
ICDAR-SROIE                   & 0.939          & 0.958      & 0.921       \\ \hline
\end{tabular}
\label{tab:agentic benckmark}
\end{table}

Integrating document classification and schema building stages creates a fully automated agentic extraction pipeline. Table \ref{tab:agentic} reports on classification confidence (linear probability) and schema evaluation scores—averaged per document type. These metrics are as critical in evaluating the agentic workflow.

{\small
\begin{table}[H]
\centering
\caption{Evaluation report on document classification and schema generation stages} 
\begin{tabular}{|l|l|l|l|l|}
\hline
\textbf{Document Type}   & \textbf{Confidence} & \textbf{Accuracy} & \textbf{Best Complexity}  \\ \hline
Invoices                     & 98.327\%         & 0.84          & 0.547         \\ \hline
Purchase Orders              & 98.196\%          & 1.00         & 0.620        \\ \hline
Receipts                     & 87.482\%          & 0.910       & 0.288        \\ \hline
Financial documents         & 98.292\%          & 0.928       & 0.434        \\ \hline
Utility Bills               & 98.5\%          & 0.903       & 0.727         \\ \hline
Salary Slips                & 1.00\%          & 1.00       & 0.488        \\ \hline
\end{tabular}
\label{tab:agentic}
\end{table}
}

Salary slips are classified with 100\% confidence, while some financial documents and utility bills are misidentified as 'unknown'. This is likely due to classification relying only on the first page, which may be a cover page in financial documents, and the low-resolution images typical of utility bills. Invoices exhibit lower classification accuracy due to false positives, as many receipts are incorrectly predicted as invoices.Schema complexity is lowest for receipts due to its simple structure, and highest for utility bills. The schema building component improves the extraction but introduces variability, as each schema is tailored per document. Documents with similar layouts but different content may produce different metadata, which is acceptable for extraction but problematic for downstream ETL processes that require uniform structures.

 \begin{figure}[H]
     \centering
     \includegraphics[width=0.8\textwidth]{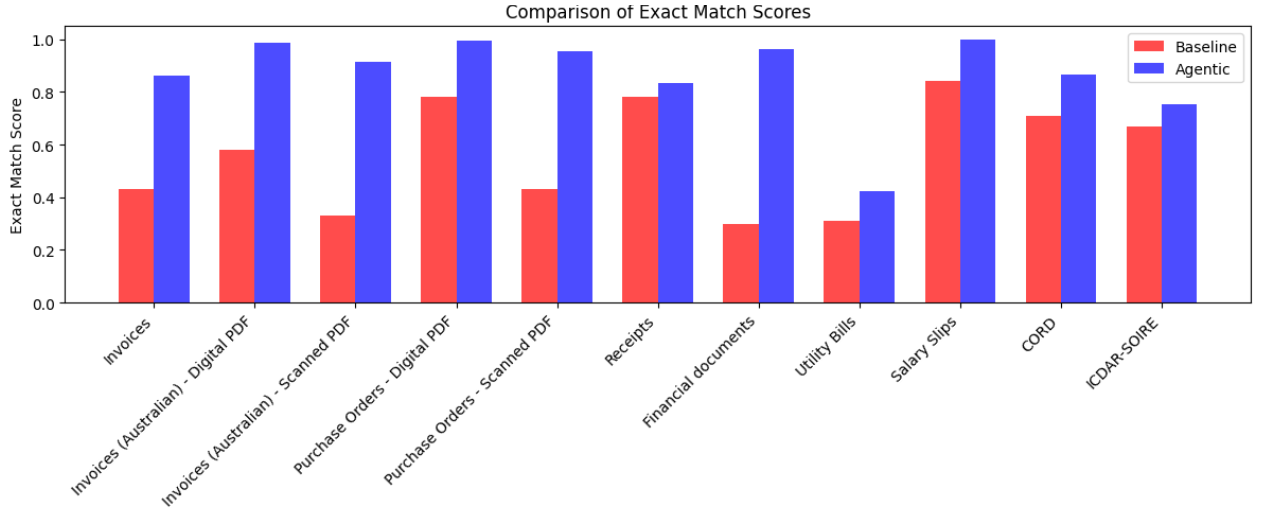}
     \caption{Baseline vs Agentic - comparison on exact matches}
     \label{fig:comparison}
 \end{figure}

As illustrated in figure \ref{fig:comparison}, comparing baseline vs agentic frameworks, where latter shows improvements across all document types with financial documents and Australian invoices reporting massive $\geq$ 0.5 point increase in exact match scores. Same is the case for all other evaluation metrics, concluding that the agentic process has shown remarkable potential to the form-like data extraction problem in an automated environment.

\section{Conclusion}

Single prompt driven applications, leveraging LLMs, have shown promising results in simple data extraction tasks. These applications are effortless, fast, and produces quality output for single-page documents. However, it faces inherent limitations in handling complex, multi-page form-like documents. Common challenges include limited context window, hallucination due to information density, and predefined prompts faltering on varying documents. This constrains the reliability and scalability of monolithic preset-prompt frameworks.

The challenges are mitigated by Agentic systems. Modular, multi-agent architectures distribute tasks and improve robustness. Systems deal inaccuracies by leveraging specialized agents for classification, splitting, schema generation and adaptive extraction. These improvements makes the process ideal for mid-size documents of good resolution. Moreover, their ability to provide confidence metrics and iterative feedback ensures higher accuracy and scalability, making them an essential evolution for form-like document data extraction.

\subsection{Limitations}

In addition to OCR reliance, we have observed that meta-prompting is sensitive to the wordings of the initial prompt. Processing speed is another challenge as agentic takes a minimum of 1 minute to extract data, while the baseline takes less than 10 seconds on average. The system caters to only a handful of document categories, and there is no automated way of handling documents classified as 'unknown'. Future studies may explore multimodality of LLMs for image data extraction, in addition to adding more learnable parameters for RL policy.

\subsection{Ethical Statements}

The usage of cloud-based LLMs to process documents containing critical information has been under ethical scrutiny since the release of OpenAI gpt-3.5. The risks include data privacy, potential security breaches, unauthorized access, and compliance with regulatory frameworks like GDPR and HIPAA.

All data used in this study is handled in accordance with the best practices of encryption and access control. The research utilizes OpenAI's enterprise API, which is designed to ensure a secure and compliant processing. Moreover, evaluation scores are reported on benchmark datasets to compare the performance of the system with existing state-of-the-art methodologies, to avoid making confidential proprietary data public.

%
%
%
\bibliographystyle{splncs04}
 \bibliography{EMAS-LNCS}
\ifdefined\isappendix
    \include{EMAS-LNCS/EMAS-LNCS-Appendix}
\fi

\end{document}